\documentclass[12pt]{article}
\usepackage{graphicx}
\title{A Time-Varying Complex Dynamical Network Model
And Its Controlled Synchronization Criteria\thanks{Accepted by
{\it IEEE Transactions on Automatic Control} on 19 May 2004. This
work was supported by the Hong Kong Research Grants Council under
the CERG grant CityU 1115/03E and the National Natural Science
Foundation of China under Grant No.60304017 and Grant
No.20336040/B06.}}
\author{Jinhu L\"u$^{1,2}$\thanks{jhlu@iss.ac.cn}
\and Guanrong Chen$^{2}$\thanks{Corresponding author e-mail:
gchen@ee.cityu.edu.hk}}
\date{{\small $^{1}$Institute of Systems Science,
Academy of Mathematics and Systems Science \\
Chinese Academy of Sciences, Beijing 100080, China\\
$^{2}$Department of Electronic Engineering, City University of
Hong Kong}}
\begin{document}
\maketitle

\begin{center}
\begin{minipage}{16cm}
{\it Abstract:} {\bf Today, complex networks have attracted
increasing attention from various fields of science and
engineering. It has been demonstrated that many complex networks
display various synchronization phenomena. In this paper, we
introduce a time-varying complex dynamical network model. We then
further investigate its synchronization phenomenon and prove
several network synchronization theorems. Especially, we show that
synchronization of such a time-varying dynamical network is
completely determined by the inner-coupling matrix, and the
eigenvalues and the corresponding eigenvectors of the coupling
configuration matrix of the network.} \\[5pt]
\end{minipage}
\end{center}

\section*{\center{I. INTRODUCTION}}

A complex network is a large set of interconnected nodes, in which
a node is a fundamental unit, which can have different meanings in
different situations, such as chemical substrates, microprocessors
or computers, schools or companies, papers, webs, even people, and
so on [1-19]. Examples of complex networks include the Internet,
the World Wide Web, food webs, electric power grids, cellular and
metabolic networks, etc. [1,2]. These large-scale complex networks
often display better cooperative or synchronous behaviors among
their constituents.

Traditionally, complex networks were studied by graph theory,
where a complex network is described by a random graph, for which
the basic theory was introduced by Erd\H{o}s and R\'{e}nyi [16].
Recently, Watts and Strogatz (WS) [4] introduced the concept of
small-world networks to describe a transition from a regular
lattice to a random graph. The WS networks exhibit a high degree
of clustering as in the regular networks and a small average
distance between two nodes as in the random networks. Moreover,
the random graph model and the WS model are both exponentially
growing networks, which are homogeneous in nature. However,
according to Barab\'asi and Albert [3], empirical results show
that many large-scale complex networks are scale-free, such as the
Internet, the WWW, and metabolic networks, among others. Notably,
a scale-free network is inhomogeneous in nature; that is, most
nodes have very few connections but a small number of particular
nodes have many connections.

In order to better understand the dynamical behaviors of various
complex networks, one may extend the existing network models from
static to dynamic by introducing dynamical elements into the
network nodes. For the resulting dynamical networks, one
significant and interesting phenomenon is the synchrony of all
dynamical nodes [5-12,15,17-21]. In fact, synchronization is a
basic motion in nature that has been studied for a long time
[5-12,17-21]. Recently, the synchronization of chaotic coupled
networks has become a focal point in the study of nonlinear
dynamics [5-12,15,17-21]. We observe that most existing work have
been devoted to completely regular networks, including lattices,
fully-connected graphs, grids, and chains, etc. [5-12,17,21],
where examples include discrete-time coupled map lattices (CML)
[5] and continuous-time cellular neural networks (CNN) [13]. A
common characteristic of these kinds of regular networks is that
their complexity originates from the nonlinear dynamics of the
nodes, but not from the network structures.

Recently, Wang and Chen [15] introduced a simple uniform
scale-free dynamical network model. Although their model reflects
the complexity from the network structure, it is only a simple
uniform dynamical network with the same coupling strength for all
connections. In reality, complex networks are more likely to have
different coupling strengths for different connections. Moreover,
real-world complex networks, particularly biological networks, are
time-varying networks. Therefore, in this paper, we attempt to
introduce a more general time-varying dynamical network model, and
further investigate its synchronization properties. Based on this
model, we prove several network synchronization theorems, showing
that synchronization of a time-varying dynamical network is
completely determined by its inner-coupling matrix (the connecting
matrix between nodes), and the eigenvalues and the corresponding
eigenvectors of its coupling configuration matrix. On the other
hand, many real-world complex networks are directed, for instance
the World Wide Web. Our new time-varying dynamical network model
can well characterize both directed and undirected dynamical
networks.

The rest of this paper is organized as follows. A new time-varying
dynamical network model is introduced in Section II. Section III
establishes several network synchronization theorems, along with
an example for illustration. Conclusions are finally given in
Section IV.

\section*{\center{II. A TIME-VARYING COMPLEX DYNAMICAL NETWORK
MODEL}}

In this section, we introduce a general time-varying complex
dynamical network model and several mathematical preliminaries.

\subsection*{A. A General Time-Varying Dynamical Network Model}

Consider a dynamical network consisting of $N$ linearly and
diffusively coupled identical nodes, with each node being an
$n-$dimensional dynamical system. The proposed general
time-varying dynamical network is described by
\begin{equation}\label{1}
  \dot{{\bf x}}_i\,=\,{\bf f}({\bf x}_i)\,
                  +\,\sum\limits^N_{\stackrel{j=1}{j\neq i}}
                  c_{ij}(t)\,{\bf A}(t)\,({\bf x}_j\,-\,{\bf x}_i),
                  \quad\quad i\,=\,1,\,2,\,\cdots,\,N\,,
\end{equation}
where ${\bf
x}_i\,=\,\left(x_{i1},\,x_{i2},\,\cdots,\,x_{in}\right)^T\in
{\textbf R}^n$ is the state variable of node $i$, ${\bf
A}(t)\,=\,\left(a_{kl}(t)\right)_{n\times n}\in
\textbf{R}^{n\times n}$ is the inner-coupling matrix of the
network at time $t$, ${\bf
C}(t)\,=\,\left(c_{ij}(t)\right)_{N\,\times\,N}$ is the coupling
configuration matrix representing the coupling strength and the
topological structure of the network at time $t$, in which
$c_{ij}(t)$ is defined as follows: if there is a connection from
node $i$ to node $j\;(j\neq i)$ at time $t$, then
$c_{ij}(t)\,\neq\,0$; otherwise, $c_{ij}(t)\,=\,0\;(j\neq i)$, and
the diagonal elements of matrix ${\bf C}(t)$ are defined by
\begin{equation}\label{2}
  c_{ii}(t)\,=\,-\sum\limits^N_{\stackrel{j=1}{j\neq i}}c_{ij}(t)\,,
  \quad\quad i\,=\,1,\,2,\,\cdots,\,N\,.
\end{equation}
The above time-varying network (\ref{1}) can be rewritten as
\begin{equation}\label{3}
    \dot{{\bf x}}_i\,=\,{\bf f}({\bf x}_i)\,+\,\sum\limits^N_{j=1}
    c_{ij}(t)\,{\bf A}(t)\,{\bf x}_j,\quad\quad i\,=\,1,\,2,\,\cdots,\,N\,.
\end{equation}
Suppose that network (\ref{3}) is connected in the sense that
there are no isolate clusters, that is, $C(t)$ is irreducible. We
assume that with their associated inner coupling matrix the
oscillators do not display Turing [20] bifurcations or other
instabilities at larger coupling values and, hence, the results in
this paper hold only for that class of oscillators and coupling
matrices.

When $\textbf{A}(t),\,\textbf{C}(t)$ are constant matrices,
network (\ref{3}) becomes a time-invariant dynamical network:
\begin{equation}\label{4}
    \dot{{\bf x}}_i\,=\,{\bf f}({\bf x}_i)\,+\,\sum\limits^N_{j=1}
    c_{ij}\,{\bf A}\,{\bf x}_j,\quad\quad i\,=\,1,\,2,\,\cdots,\,N\,.
\end{equation}

Obviously, the simple uniform dynamical network of Wang and Chen
[15],
\begin{equation}\label{5}
    \dot{{\bf x}}_i\,=\,{\bf f}({\bf x}_i)\,+\,c\,\sum\limits^N_{j=1}
    c_{ij}\,{\bf A}\,{\bf x}_j,\quad\quad i\,=\,1,\,2,\,\cdots,\,N\,,
\end{equation}
where ${\bf C}$ is a $0-1$ matrix and ${\bf A}$ is a $0-1$
diagonal matrix, is a special case of network (\ref{4}).

\subsection*{B. Mathematical Preliminaries}

\noindent{\bf Lemma 1:} Suppose that an irreducible constant
matrix ${\bf C}\,=\,\left(c_{ij}\right)_{N\times N}$ satisfies
condition (\ref{2}). Then:
\begin{itemize}
\item[\rm{(i)}] $0$ is an eigenvalue of matrix ${\bf C}$,
associated with eigenvector $\left(1,\,1,\,\cdots,\,1\right)^T$.
\item[\rm{(ii)}] If $c_{ij}\,\geq\,0$ for all $1\,\leq\,
i,\,j\,\leq\,N\;(j\,\neq\,i)$, then the real parts of all
eigenvalues of ${\bf C}$ are less than or equal to $0$, and all
possible eigenvalues with zero real part are the real eigenvalue
$0$. In fact, $0$ is its eigenvalue of multiplicity $1$.
\end{itemize}
\medskip

The proof is omitted here since it can easily be deduced from the
Gerschgorin's circle theorem and the Perron-Frobenius theorem
[21,22].

\medskip
\noindent{\bf Definition 1:} Let ${\bf x}_i(t\,,\,{\bf
X}_0 )\;(i\,=\,1,\,2,\,\cdots,\,N)$ be a solution of the
nonautonomous dynamical network
\begin{equation}\label{6}
\dot{{\bf x}}_i\,=\,{\bf f}({\bf x}_i)\,+\,{\bf g}_i(t,\,{\bf
x}_1,\,{\bf x}_2,\,\cdots,\,{\bf x}_N),\quad\quad
i\,=\,1,\,2,\,\cdots,\,N\,,
\end{equation}
where ${\bf X}_0\,=\,\left(({\bf x}_1^0)^T,\,\cdots,\,({\bf
x}_N^0)^T\right)^T\in\,\textbf{R}^{nN}$, ${\bf
f}\,:\,D\rightarrow{\textbf R}^n$ and ${\bf g}_i\,:\,D\times
\cdots\times D\rightarrow{\textbf
R}^n\,(i\,=\,1,\,2,\,\cdots,\,N)$ are continuously differentiable
with $D\,\subseteq\,{\textbf R}^n$, and ${\bf g}_i(t,\,{\bf
x},\,{\bf x},\,\cdots,\,{\bf x})\,=\,0$ for all $t$. If there is a
nonempty open subset $E\subseteq D$, with ${\bf x}_i^0\in\,E\,(i\,
=\,1,\,2,\,\cdots,\,N)$, such that ${\bf x}_i(t\,,\,{\bf
X}_0)\,\in\,D$ for all $t\,\geq\,0$, $i\,=\,1,\,2,\,\cdots,\,N$,
and
\begin{equation}\label{7}
\lim\limits_{t\rightarrow\infty}\left\|{\bf x}_i(t\,,\,{\bf
X}_0)-{\bf s}(t\,,\,{\bf x}_0)\right\|_2\,=\,{\bf 0}
\quad\quad\mbox{for}\quad 1\leq\,i\,\leq N\,,
\end{equation}
where ${\bf s}(t\,,\,{\bf x}_0)$ is a solution of the system
$\dot{{\bf x}}\,=\,{\bf f}({\bf x})$ with ${\bf x}_0\in\,D$, then
the dynamical network (\ref{6}) is said to realize {\it
synchronization} and $E\times\cdots\times E$ is called the {\it
region of synchrony} for network (\ref{6}). Moreover, ${\bf
X}(t,{X_0})\,=\,\left({\bf x}_1^T(t,{\bf X}_0),\,{\bf
x}_2^T(t,{\bf X}_0),\,\cdots,\,{\bf x}_N^T(t,{\bf X}_0)\right)^T$
is called the {\it synchronous solution} of network (\ref{6}), if
${\bf x}_i(t,{\bf X}_0)={\bf x}_j(t,{\bf X}_0)$ for all $t\geq0$
and $1\leq i,\,j\leq N$.
\medskip

\noindent{\bf Definition 2:} [23] Let $\Gamma\,=\,\{{\bf
s}(t)\;|\; 0\leq t\,<\,T\}$ denote the set of $T-$periodic
solutions of system $\dot{{\bf x}}\,=\,{\bf f}({\bf x})$ in
$\textbf{R}^n$. A $T-$periodic solution ${\bf s}(t)$ is said to be
{\it orbitally stable}, if for each $\varepsilon\,>\,0$ there
exists a $\delta\,>\,0$ such that every solution ${\bf x}(t)$ of
$\dot{{\bf x}}\,=\,{\bf f}({\bf x})$, whose distance from $\Gamma$
is less than $\delta$ at $t\,=\,0$, will remain within a distance
less than $\varepsilon$ from $\Gamma$ for all $t\,\geq\,0$. Such
an ${\bf s}(t)$ is said to be {\it orbitally asymptotically
stable} if, in addition, the distance of ${\bf x}(t)$ from
$\Gamma$ tends to zero as $t\,\rightarrow\,\infty$. Moreover, if
there exist positive constants $\alpha,\,\beta$ and a real
constant $h$ such that $\left\|{\bf x}(t-h)\,-\,{\bf
s}(t)\right\|\,\leq\,\alpha\,e^{-\beta t}$ for $t\,\geq\,0$, then
${\bf s}(t)$ is said to be {\it orbitally asymptotically stable
with an asymptotic phase}.
\medskip

For network (\ref{3}), diffusive coupling condition (\ref{2})
ensures that $({\bf x}_1^T(t,{\bf X}_0),{\bf x}_2^T(t,{\bf
X}_0),\cdots,{\bf x}_N^T(t,{\bf X}_0))^T$ with ${\bf
x}_i(t\,,\,{\bf X}_0)\,=\,{\bf s}(t\,,\,{\bf
x}_0)\;(1\,\leq\,i\,\leq\,N)$ is a synchronous solution of network
(\ref{3}), where ${\bf X}_0\,=\,(({\bf x}_1^0)^T,\,\cdots,({\bf
x}_N^0)^T)^T$, ${\bf x}^0_1\,=\,\cdots\,=\,{\bf x}^0_N\,=\,{\bf
x}_0\,\in\,D$, and ${\bf s}(t\,,\,{\bf x}_0)$ is a solution of
system $\dot{{\bf x}}\,=\,{\bf f}({\bf x})$. It is noted that
${\bf s}(t\,,\,{\bf x}_0)$ can generate an equilibrium point, a
periodic orbit, an aperiodic orbit, or a chaotic orbit in the
phase space. In the following, we only study the case that ${\bf
s}(t\,,\,{\bf x}_0)$ {\it cannot} generate a chaotic orbit, since
we have already investigated the case of chaos synchronization in
general time-varying dynamical networks in [18].

Note that non-chaotic synchronization and chaos synchronization
are different in essence. It is very important to point out that
since we do not assume the stability of ${\bf s}(t\,,\,{\bf x}_0)$
for system $\dot{{\bf x}}\,=\,{\bf f}({\bf x})$, the stability of
the synchronous solution ${\bf S}(t)=({\bf s}^T(t,\,{\bf
x}_0),\,{\bf s}^T(t,\,{\bf x}_0),\,\cdots,{\bf s}^T(t,\,{\bf
x}_0))^T$ of network (\ref{3}) is equivalent to the stability of
the {\it error vector} $\left({\bf \eta}_1^T(t),\,{\bf
\eta}_2^T(t),\,\cdots,\,{\bf \eta}_N^T(t)\right)^T$ about its zero
solution, where ${\bf \eta}_i(t)\,=\,{\bf x}_i(t)\,-\,{\bf
s}(t\,,\,{\bf x}_0)\;(i\,=\,1,\,2,\,\cdots,\,N)$. However, it is
quite different from the chaos synchronization case, since a
chaotic attractor is an attracting invariant set and a chaotic
orbit is not stable in Lyapunov sense, so the stability of the
chaotic synchronous state ${\bf x}_1(t)\,=\,{\bf
x}_2(t)\,=\,\cdots\,=\,{\bf x}_N(t)$ is equivalent to the
stability of the {\it transverse errors} $\left({\bf
\eta}_2^T(t),\, {\bf \eta}_3^T(t),\,\cdots,\,{\bf
\eta}_N^T(t)\right)^T$ of the synchronous manifold of network
(\ref{3}), where ${\bf \eta}_i(t)\,=\,{\bf x}_i(t)\,-\,{\bf
s}(t\,,\,{\bf x}_0)\; (i\,=\,2,\,3,\,\cdots,\,N)$ and ${\bf
x}_1(t)\,=\,{\bf s}(t\,,\,{\bf x}_0)$ is the reference direction
of the synchronous manifold [18,19].

\section*{\center{III. SEVERAL NETWORK SYNCHRONIZATION CRITERIA}}

In this section, we establish several network synchronization
criteria for both the time-varying network (\ref{3}) and the
time-invariant network (\ref{4}).

\subsection*{A. Periodic Orbits Synchronization of Time-Invariant
Dynamical Networks}

Here, we consider periodic orbits synchronization of the
time-invariant network (\ref{4}). Assume that ${\bf s}(t\,,\,{\bf
x}_0)$ is a nonconstant $T-$periodic solution of an individual
node $\dot{{\bf x}}\,=\,{\bf f}({\bf x})$, denoted ${\bf s}(t)$.

Now, let
\begin{equation}\label{8}
  {\bf \eta}_i(t)\,=\,{\bf x}_i(t)\,-\,{\bf s}(t),\quad\quad
                    i\,=\,1,\,2,\,\cdots,\,N\,,
\end{equation}
and substitute (\ref{8}) into network (\ref{4}) to get
\begin{equation}\label{9}
\dot{{\bf \eta}}_i(t)\,=\,{\bf f}({\bf s}(t)\,+\,{\bf
                        \eta}_i(t))\,-\,{\bf f}({\bf s}(t))\,
                        +\,\sum\limits^N_{j=1}c_{ij}\,{\bf A}\,
                        {\bf\eta}_j(t)\,,\quad 1\leq i\leq N\,.
\end{equation}
Also, denote
$$
{\bf X}(t)\,=\,\left(\begin{array}{c}
  {\bf x}_1(t) \\
  \vdots \\
  {\bf x}_N(t) \\
\end{array}\right)\,,\quad\quad\bar{{\bf \eta}}(t)\,
                     =\,\left(\begin{array}{c}
  {\bf \eta}_{1}(t) \\
  \vdots \\
  {\bf \eta}_{N}(t) \\
\end{array}\right)\,,\quad\quad{\bf S}(t)\,=\,\left(\begin{array}{c}
  {\bf s}(t) \\
  \vdots \\
  {\bf s}(t) \\
\end{array}\right)\,\in\,\textbf{R}^{nN}\,.
$$
Then, (\ref{9}) can be rewritten as
\begin{equation}\label{10}
\dot{\bar{{\bf\eta}}}(t)\,=\,{\bf F}\left(t,\,\bar{{\bf
                           \eta}}(t)\right)\,,
\end{equation}
and the Jacobian of ${\bf F}(t,\,\bar{{\bf \eta}})$ at $\bar{{\bf
\eta}}\,=\,0$ is
$$
  D{\bf F}(t,\,{\bf 0})\,=\,\left(\begin{array}{cccc}
  D{\bf f}({\bf s}(t))\,+\,c_{11}\,{\bf A} & c_{12}\,{\bf A}
  & \cdots & c_{1N}\,{\bf A} \\
  c_{21}\,{\bf A} & D{\bf f}({\bf s}(t))\,+\,c_{22}\,{\bf A}
  & \cdots & c_{2N}\,{\bf A} \\
  \vdots & \vdots & \ddots & \vdots \\
  c_{N1}\,{\bf A} & c_{N3}\,{\bf A} & \cdots
  & D{\bf f}({\bf s}(t))\,+\,c_{NN}\,{\bf A} \\
\end{array}\right)\,.
$$

Obviously, ${\bf S}(t)\,=\,\left({\bf s}^T(t),\,{\bf
s}^T(t),\,\cdots,\,{\bf s}^T(t)\right)^T$ is a synchronous
periodic solution of network (\ref{4}). It is interesting and also
important to find out whether or not this synchronous periodic
solution ${\bf S}(t)$ is also orbitally asymptotically stable in
network (\ref{4}) if ${\bf s}(t)$ is orbitally asymptotically
stable in an individual node described by $\dot{{\bf x}}\,=\,{\bf
f}({\bf x})$. The following theorem gives a positive answer to
this question.
\medskip

\noindent{\bf Definition 3:} Suppose that ${\bf s}(t)$ is a
periodic solution of system $\dot{{\bf x}}\,=\,{\bf f}({\bf x})$.
Let $\gamma_1\,=\,1,\,\gamma_2,\,\cdots,\,\gamma_n$ be the Floquet
multipliers of the variational equation of ${\bf s}(t)$,
$\dot{{\bf y}}\,=\,{\bf A}(t)\,{\bf y}$, where ${\bf
A}(t)\,=\,D{\bf f}({\bf s}(t))$ is the Jacobian of ${\bf f}$
evaluated at ${\bf s}(t)$. Then the periodic solution ${\bf s}(t)$
is said to be a {\it hyperbolic periodic solution} if
$|\gamma_j|\,\neq\,1$ for $2\,\leq\,j\,\leq n$. Moreover, ${\bf
S}(t)$ is said to be a {\it hyperbolic synchronous periodic
solution} of network (\ref{4}) if all Floquet multipliers of the
variational equation of ${\bf S}(t)$ have absolute values less
than $1$ except one multiplier which equals 1.
\bigskip

\noindent{\bf Theorem 1:} Suppose that ${\bf s}(t)$ is a
hyperbolic periodic solution of an individual node $\dot{{\bf
x}}\,=\,{\bf f}({\bf x})$, and is orbitally asymptotically stable
with an asymptotic phase. Suppose also that the coupling
configuration matrix ${\bf C}\,=\, \left(c_{ij}\right)_{N\times
N}$ can be diagonalized. Then, ${\bf S}(t)$ is a hyperbolic
synchronous periodic solution of network (\ref{4}), and is
orbitally asymptotically stable with an asymptotic phase, if and
only if the linear time-varying systems
\begin{equation}\label{11}
  \dot{{\bf w}}\,=\,[D{\bf f}({\bf s}(t))\,+\,\lambda_k\,{\bf A}]\,
                  {\bf w}\,,\quad\quad k\,=\,2,\,\cdots,\,N\,,
\end{equation}
are asymptotic stable about their zero solutions.
\medbreak

\noindent{\bf Proof.} It is well known that the stability of a
periodic solution of an autonomous system is completely determined
by its Floquet multipliers [23]. We linearize network (\ref{4}) at
${\bf X}(t)\,=\,{\bf S}(t)$ and get
\begin{equation}\label{12}
\dot{\bar{{\bf Y}}}(t)\,=\,D{\bf F}(t,\,{\bf 0})\,\bar{{\bf\,Y}}(t)\,,
\end{equation}
where $D{\bf F}(t,\,{\bf 0})$ is defined in (\ref{10}) and
$\bar{{\bf Y}}(t)=\left({\bf y}_1^T(t),\,\cdots,\,{\bf
y}_N^T(t)\right)^T\in\textbf{R}^{n N}$. According to the stability
theory of periodic orbits [23], ${\bf S}(t)$ is a hyperbolic
synchronous periodic solution of network (\ref{4}), and is
orbitally asymptotically stable with an asymptotic phase, if and
only if all Floquet multipliers of system (\ref{12}) have absolute
values less than $1$ except one multiplier which equals 1.

From (\ref{12}), we have
$$
\begin{array}{rcl}
\dot{{\bf y}}_i(t)
   & = & D{\bf f}({\bf s}(t))\,{\bf y}_i(t)\,+\,\sum\limits^N_{j=1}c_{ij}
         \,{\bf A}\,{\bf y}_j(t)\\
   & = & D{\bf f}({\bf s}(t))\,{\bf y}_i(t)\,+\,{\bf A}\,\left({\bf y}_1(t),
         \,\cdots,\,{\bf y}_N(t)\right)\left(c_{i1},\,\cdots,\,c_{iN}\right)^T,
         \quad i\,=\,1,\,2,\,\cdots,\,N\,,
\end{array}
$$
where ${\bf Y}(t)\,=\,\left({\bf y}_1(t),\,\cdots,\,{\bf
y}_N(t)\right)\in\textbf{R}^{n\times N}$. That is,
\begin{equation}\label{13}
  \dot{{\bf Y}}(t)\,=\,D{\bf f}({\bf s}(t))\,{\bf Y}(t)\,
                     +\,{\bf A}\,{\bf Y}(t)\,{\bf C}^T\,.
\end{equation}
Obviously, systems (\ref{12}) and (\ref{13}) have the same Floquet
multipliers, since they are different only in form.

Since ${\bf C}$ can be diagonalized, namely, it has all real
eigenvalues of multiplicity one, there exists an nonsingular
matrix, ${\bf \Phi}\,=\,\left({\bf \phi}_1,\,\cdots,\,{\bf
\phi}_N\right)$, such that
$$
  {\bf C}^T\,{\bf \Phi}\,=\,{\bf \Phi}\,{\bf \Lambda}\,,
$$
where ${\bf
\Lambda}\,=\,\mbox{diag}\{\lambda_1,\,\lambda_2,\,\cdots,\,\lambda_N\}$,
and $\lambda_1,\,\cdots,\,\lambda_N$ are the eigenvalues of ${\bf
C}$. Let
$$
  {\bf Y}\,(t)\,=\,{\bf V}(t)\,{\bf \Phi}^{-1}\,.
$$
From (\ref{13}), the matrix vector ${\bf V}(t)\,=\,\left({\bf
v}_1(t),\,\cdots,\,{\bf v}_N(t)\right)\,\in\,\textbf{R}^{n\times
N}$ satisfies the following equation:
$$
\dot{{\bf V}}(t)\,=\,D{\bf f}({\bf s}(t))\,{\bf V}(t)\,+\,{\bf
A}\,{\bf V}(t)\,{\bf \Lambda}\,,
$$
namely,
\begin{equation}\label{14}
  \dot{{\bf v}}_k(t)\,=\,[D{\bf f}({\bf s}(t))\,+\,\lambda_k\,{\bf A}]
  \,{\bf v}_k(t)\,,\quad k\,=\,1,\,2,\,\cdots,\,N\,.
\end{equation}
Thus, we have changed the stability problem of the synchronous
periodic solution ${\bf S}(t)$ into the stability problem of $N$
independent $n-$dimensional linear systems (\ref{14}), which has
the same form as (\ref{11}).

It follows from Lemma 1 that one eigenvalue of ${\bf C}$ satisfies
$\lambda_1\,=\,0$ and the corresponding linear system in
(\ref{14}) is
\begin{equation}\label{15}
  \dot{{\bf v}}_1(t)\,=\,D{\bf f}({\bf s}(t))\,{\bf v}_1(t)\,.
\end{equation}
Obviously, system (\ref{15}) is the corresponding linearized
system of an individual node $\dot{{\bf x}}\,=\,{\bf f}({\bf x})$
at ${\bf x}\,=\,{\bf s}(t)$. When $k\,>\,1$, systems (\ref{14})
become systems (\ref{11}).

Let $\gamma_{ij}\;(i\,=\,1,\,2,\,\cdots,\,N;\;j\,=\,1,\,2,\,
\cdots,\,n)$ be the Floquet multipliers of the $N$ independent
$n-$dimensional linear time-varying systems (\ref{14}). Obviously,
the Floquet multipliers of system (\ref{12}) (or (\ref{13})) are
also given by $\gamma_{ij}\;(i\,=\,1,\,2,\,\cdots,\,N;\;j\,=\,
1,\,2,\,\cdots,\,n)$.

According to the stability theory of periodic orbits [23], ${\bf
s}(t)$ is a hyperbolic periodic solution of an individual node
$\dot{{\bf x}}\,=\,{\bf f}({\bf x})$, and is orbitally
asymptotically stable with an asymptotic phase, if and only if all
Floquet multipliers of (\ref{15}) satisfy $|\gamma_{1j}|\,<\,1$
for $j\,=\,2,\,\cdots,\,n$, except one multiplier $\gamma_{11}$
which equals $1$. Moreover, the linear systems (\ref{11}) are
asymptotically stable if and only if all Floquet multipliers of
systems (\ref{11}) satisfy $|\gamma_{ij}|\,<\,1$ for
$i\,=\,2,\,\cdots,\,N$ and $j\,=\,1,\,2,\,\cdots,\,n$. Therefore,
${\bf S}(t)$ is a hyperbolic synchronous periodic solution of
dynamical network (\ref{4}), and is orbitally asymptotically
stable with an asymptotic phase, if and only if all Floquet
multipliers of the linear systems (\ref{14}) satisfy
$|\gamma_{1j}|\,<\,1$ for $j\,=\,2,\,\cdots,\,n$ and
$|\gamma_{ij}|\,<\,1$ for
$i\,=\,2,\,\cdots,\,N;\;j\,=\,1,\,2,\,\cdots,\,n$, except
$\gamma_{11}\,=\,1$.

The proof is thus completed.
\medskip

Note that the result of Theorem 1 can not be extended to the
time-varying dynamical network (\ref{3}), which is a nonautonomous
system. Network (\ref{3}) is discussed in the next subsection.

\subsection*{B. Stable Orbits Synchronization of Time-Varying
Dynamical Networks}

Here, assume that ${\bf s}(t\,,\,{\bf x}_0)$ is an exponentially
stable solution of $\dot{{\bf x}}\,=\,{\bf f}({\bf x})$, simply
denoted as ${\bf s}(t)$. Let $\mu\,({\bf A})$ be the maximum
eigenvalue of $\frac{1}{2}\left({\bf A}^T\,+\,{\bf A}\right)$. We
provide two network synchronization theorems for the time-varying
dynamical network (\ref{3}) in this subsection.

Substituting (\ref{8}) into the time-varying network (\ref{3})
yields
\begin{equation}\label{16}
    \dot{{\bf \eta}}_i(t)\,=\,{\bf f}({\bf s}(t)\,
    +\,{\bf \eta}_i(t))\,-\,{\bf f}({\bf s}(t))\,
    +\,\sum\limits^N_{j=1}c_{ij}(t)\,{\bf A}(t)
    \,{\bf \eta}_j(t)\,,\quad 1\leq i\leq N\,.
\end{equation}
Similar to (\ref{10}), system (\ref{16}) can be reformulated as
\begin{equation}\label{17}
    \dot{\bar{{\bf \eta}}}(t)\,=\,\bar{{\bf F}}
    \left(t,\,\bar{{\bf \eta}}(t)\right)\,,
\end{equation}
and the Jacobian of $\bar{{\bf F}}(t,\,\bar{{\bf \eta}})$ at
$\bar{{\bf \eta}}\,=\,{\bf 0}$ is
$$
D\bar{{\bf F}}(t,{\bf 0})=\left(\begin{array}{cccc}
  D{\bf f}({\bf s}(t))+c_{11}(t){\bf A}(t) & c_{12}(t){\bf A}(t)
  & \cdots & c_{1N}(t){\bf A}(t) \\
  c_{21}(t){\bf A}(t) & D{\bf f}({\bf s}(t))+c_{22}(t){\bf A}(t)
  & \cdots & c_{2N}(t){\bf A}(t) \\
  \vdots & \vdots & \ddots & \vdots \\
  c_{N1}(t){\bf A}(t) & c_{N2}(t){\bf A}(t)
  & \cdots & D{\bf f}({\bf s}(t))+c_{NN}(t){\bf A}(t) \\
\end{array}\right).
$$
\bigskip

\noindent{\bf Theorem 2:} Suppose that $\bar{{\bf
F}}:\,\bar{\Omega}\rightarrow\,\textbf{R}^{nN}$ is continuously
differentiable on $\bar{\Omega}\,=\,\{\bar{{\bf
\eta}}\in\,\textbf{R}^{nN}\,|\|\bar{{\bf \eta}}\|_2\,<\,r\}$. The
synchronous solution ${\bf S}(t)$ is uniformly exponentially
stable in the dynamical network (\ref{3}) if there exist two
symmetric positive definite matrices, ${\bf P},\;{\bf
Q}\,\in\,\textbf{R}^{nN\times nN}$, such that
$$
{\bf P}\,\left(D\bar{{\bf F}}(t,\,{\bf
0})\right)\,+\,\left(D\bar{{\bf F}}(t,\,{\bf 0})\right)^T\,{\bf
P}\,\leq\,-\,{\bf Q}\,\leq\,-\,c_1\,{\bf I}\,,
$$
where $c_1\,>\,0$, and
$$
\left({\bf \Gamma}(t,\,{\bf y})\,-\,{\bf \Gamma}(t,\,{\bf
S}(t))\right)^T\,{\bf P}\,+\,{\bf P}\,\left({\bf \Gamma}(t,\,{\bf
y})\,-\,{\bf \Gamma}(t,\,{\bf S}(t))\right)\,
     \leq\,c_2\,{\bf I}\,<c_1\,{\bf I}\,,
$$
in which ${\bf \Gamma}(t,\,{\bf y}(t))\,=\,\mbox{diag}\{D{\bf
f}({\bf y}_1(t)),\,\cdots,\,D{\bf f}({\bf y}_{N}(t))\}$, ${\bf
y}(t)\,=\,\left({\bf y}^T_1(t),\,{\bf y}^T_2(t),\,\cdots,\,{\bf
y}^T_{N}(t)\right)^T$, and ${\bf y}-{\bf
S}(t)\,\in\,\bar{\Omega}$.
\medskip

\noindent{\bf Proof.} Since $\bar{{\bf F}}$ is continuously
differentiable on $\bar{\Omega}$, from (\ref{16})-(\ref{17}), we
have
$$
\begin{array}{rcl}
\dot{\bar{{\bf \eta}}}(t) & = & \left(\begin{array}{c}
        {\bf f}({\bf \eta}_1(t)\,+\,{\bf s}(t))-{\bf f}({\bf s}(t)) \\
        \vdots \\
        {\bf f}({\bf \eta}_N(t)\,+\,{\bf s}(t))-{\bf f}({\bf s}(t)) \\
        \end{array}\right)\,+\,\left(\begin{array}{ccc}
        c_{11}(t)\,{\bf A}(t) & \cdots & c_{1N}(t)\,{\bf A}(t) \\
        \vdots & \ddots & \vdots \\
        c_{N1}(t)\,{\bf A}(t) & \cdots & c_{NN}(t)\,{\bf A}(t) \\
        \end{array}\right)\,\bar{{\bf \eta}}(t)  \\
        & = &  \left(\begin{array}{c}
        D{\bf f}({\bf y}_1(t,\,{\bf \eta}_1(t)))\,{\bf \eta}_1(t) \\
        \vdots \\
        D{\bf f}({\bf y}_N(t,\,{\bf \eta}_N(t)))\,{\bf \eta}_N(t) \\
        \end{array}\right)\,+\,\left(\begin{array}{ccc}
        c_{11}(t)\,{\bf A}(t) & \cdots & c_{1N}(t)\,{\bf A}(t) \\
        \vdots & \ddots & \vdots \\
        c_{N1}(t)\,{\bf A}(t) & \cdots & c_{NN}(t)\,{\bf A}(t) \\
        \end{array}\right)\,\bar{{\bf \eta}}(t)  \\
        & = &  \left(\begin{array}{ccc}
        D{\bf f}({\bf y}_1(t,\,{\bf \eta}_1(t)))\,-\,D{\bf f}({\bf s}(t))
        & 0 & 0 \\
        0 & \ddots & 0 \\
        0 & 0 & D{\bf f}({\bf y}_N(t,\,{\bf \eta}_N(t)))\,
                -\,D{\bf f}({\bf s}(t)) \\
        \end{array}\right)\,\bar{{\bf \eta}}(t) \nonumber \\
        &   &  +\,D{\bar{\bf F}}(t,\,{\bf 0})\,\bar{{\bf \eta}}(t)  \\
        & = &  \left[{\bf \Gamma}(t,\,{\bf y}(t))\,
               -\,{\bf \Gamma}(t,\,{\bf S}(t))\,
        +\,D{\bar{\bf F}}(t,{\bf 0})\right]\,\bar{{\bf \eta}}(t)\,.
\end{array}
$$
Define a Lyapunov function, using the vector $\bar{{\bf
\eta}}(t)$, by
$$
{\bf V}\,(t)\,=\,\bar{{\bf \eta}}(t)^T\,{\bf P}\,\bar{{\bf
\eta}}(t)\,,
$$
and differentiate ${\bf V}(t)$ with respect to time $t$ to get
$$
\begin{array}{rcl}
  \dot{{\bf V}}(t)
  & = & \dot{\bar{{\bf \eta}}}(t)^T\,{\bf P}\,\bar{{\bf \eta}}(t)\,
  +\,\bar{{\bf \eta}}(t)^T\,{\bf P}\,\dot{\bar{{\bf \eta}}}(t)\, \\
  & = & \bar{{\bf \eta}}(t)^T\,\left[\left(D{\bar{\bf F}}
  (t,\,{\bf 0})\right)^T\,{\bf P}\,+\,{\bf P}\,\left(D{\bar{\bf F}}
  (t,\,{\bf 0})\right)\right]\,\bar{{\bf \eta}}(t) \\
  &   & +\;\bar{{\bf \eta}}(t)^T\,\left[\left({\bf \Gamma}(t,\,{\bf y})\,
  -\,{\bf \Gamma}(t,\,{\bf S}(t))\right)^T\,{\bf P}\,
  +\,{\bf P}\,\left({\bf \Gamma}(t,\,{\bf y})\,
  -\,{\bf \Gamma}(t,\,{\bf S}(t))\right)\right]\,\bar{{\bf \eta}}(t) \\
  & \leq & -\,\bar{{\bf \eta}}(t)^T\,{\bf Q}\,\bar{{\bf \eta}}(t)\,
  +\,c_2\,\bar{{\bf \eta}}(t)^T\,{\bf I}\,\bar{{\bf \eta}}(t) \\
  & \leq & \left(c_2\,-\,c_1\right)\,\bar{{\bf \eta}}(t)^T\,
  \bar{{\bf \eta}}(t)\,<\,0\,.
\end{array}
$$
From the Lyapunov stability theory, the synchronization errors
$\bar{{\bf \eta}}(t)$ will uniformly exponentially converge to
zero. That is, the synchronous state ${\bf S}(t)$ is uniformly
exponentially stable for the dynamical network (\ref{3}). The
proof is thus completed.
\medskip

Next, we consider a class of time-varying dynamical networks whose
configuration matrix satisfies a special property.
\medskip

\noindent{\bf Assumption 1:} Let
$\lambda_1(t),\,\lambda_2(t),\,\cdots,\,\lambda_N(t)$ be the
eigenvalues of ${\bf C}(t)$. $\exists\,t_0\,\geq\,0$, for any
$\lambda_i(t)\;(1\leq\,i\,\leq N)$, either $\lambda_i(t)\neq0$ for
all $t\,>\,t_0$, or $\lambda_i(t)\equiv0$ for all $t\,>\,t_0$.
\medskip

For any fixed $t_1\,>\,t_0$, ${\bf C}(t_1)$ is a constant matrix.
According to Lemma 1, there exists a unique $\lambda_i(t)$
satisfying $\lambda_i(t_1)=0$ and $\lambda_j(t_1)\,<\,0$ for
$j\neq i$. From Assumption 1, the eigenvalues of ${\bf C}(t)$
satisfy $\lambda_i(t)=0$ for all $t\,>\,t_0$ and
$\lambda_j(t)\,<\,0$ for all $t\,>\,t_0$, $j\neq i$. Let
$\lambda_i(t)$ be $\lambda_1(t)$.
\medskip

\noindent{\bf Theorem 3:} Let ${\bf x}\,=\,{\bf s}(t)$ be an
exponentially stable solution of nonlinear system $\dot{{\bf
x}}\,=\,{\bf f}({\bf x})$, where ${\bf f}:\,\Omega\rightarrow\,
\textbf{R}^n$ is continuously differentiable, $\Omega\,=\,\{{\bf
x}\in\,\textbf{R}^n\,|\|{\bf x}\,-\,{\bf s}(t)\|_2\,<\,r\}$.
Suppose that the Jacobian $D\bar{{\bf F}}(t,\,\bar{{\bf \eta}})$
is bounded and Lipschitz on $\bar{\Omega}\,=\,\{\bar{{\bf
\eta}}\in\,\textbf{R}^{nN}\,|\|\bar{{\bf \eta}}\|_2\,<\,r\}$,
uniformly in $t$. Suppose also that Assumption 1 holds and there
exists a real matrix, ${\bf \Phi}(t)$, nonsingular for all $t$,
such that ${\bf \Phi}^{-1}(t)\,\left({\bf C}(t)\right)^T\,{\bf
\Phi}(t)\,=\,\mbox{diag}\{\lambda_1(t),\,\lambda_2(t),\,\cdots,\,
\lambda_N(t)\}$ and $\dot{{\bf \Phi}}^{-1} (t)\,{\bf
\Phi}(t)\,=\,\mbox{diag}\{\beta_1(t),\,\beta_2(t),\,\cdots,\,\beta_N(t)\}$.
Then, the synchronous solution ${\bf S}(t)$ is exponentially
stable in dynamical network (\ref{3}) if and only if the linear
time-varying systems
\begin{equation}\label{18}
\dot{{\bf w}}\,=\,[D{\bf f}({\bf s}(t))\,+\,\lambda_k(t)\,{\bf
                A}(t)\,-\,\beta_k(t)\,{\bf I}_n]\,{\bf
                w}\,,\quad\quad k\,=\,2,\,\cdots,\,N\,,
\end{equation}
are exponentially stable about their zero solutions.
\medskip

\noindent{\bf Proof.} Let
\begin{equation}\label{19}
  {\bf x}(t)\,=\,{\bf z}(t)\,+\,{\bf s}(t)\,,
\end{equation}
and substitute (\ref{19}) into $\dot{{\bf x}}\,=\,{\bf f}({\bf
x})$, so as to obtain
\begin{equation}\label{20}
\dot{{\bf z}}\,=\,D{\bf f}({\bf s}(t))\,{\bf z}\,+\,{\bf
h}(t,\,{\bf z})\,,
\end{equation}
where ${\bf h}(t,\,{\bf z})\,=\,{\bf f}({\bf z}\,+\,{\bf
s}(t))\,-\,{\bf f}({\bf s}(t))\,-\,D{\bf f}({\bf s}(t)){\bf z}$.

Since $D{\bar{\bf F}}(t,\,\bar{{\bf \eta}})$ is bounded and
Lipschitz on $\bar{\Omega}$, uniformly in $t$, $D{\bf f}({\bf x})$
is also bounded and Lipschitz on $\Omega$, uniformly in $t$. That
is, the Jacobian of system (\ref{20}), i.e., $\frac{\partial {\bf
f}({\bf z}\,+\,{\bf s}(t))}{\partial{\bf z}}\,=\,
\frac{\partial{\bf f}({\bf z}\,+\,{\bf s}(t))}{\partial({\bf
z}\,+\,{\bf s}(t))}\,=\,D{\bf f}({\bf x})$, is bounded and
Lipschitz on $\Omega$, uniformly in $t$. According to the Lyapunov
converse theorem [24], the origin is an exponentially stable
equilibrium point for the nonlinear system (\ref{20}) if and only
if it is an exponentially stable equilibrium point for the
corresponding linear time-varying system
\begin{equation}\label{21}
  \dot{{\bf y}}\,=\,D{\bf f}({\bf s}(t))\,{\bf y}\,.
\end{equation}

Since network (\ref{3}) can be rewritten in the form of system
(\ref{17}), we linearize network (\ref{3}) at ${\bf X}(t)={\bf
S}(t)$ and get
\begin{equation}\label{22}
    \dot{\bar{{\bf Y}}}\,=\,D{\bar{\bf F}}\left(t,\,{\bf 0}\right)
                          \,\bar{{\bf Y}}\,,
\end{equation}
where $\bar{{\bf Y}}(t)=\left({\bf y}_1^T(t),\,\cdots,\,{\bf
y}_N^T(t)\right)^T\in\textbf{R}^{n N}$.

Since $D{\bar{\bf F}}(t,\,\bar{{\bf \eta}})$ is bounded and
Lipschitz on $\bar{\Omega}$, uniformly in $t$, according to the
Lyapunov converse theorem [24], the origin is an exponentially
stable equilibrium point for the nonlinear system (\ref{17}) if
and only if it is an exponentially stable equilibrium point for
the corresponding linear time-varying system (\ref{22}). Note that
the exponential stability of the zero solution of system
(\ref{22}) means that $\bar{{\bf Y}}(t)\rightarrow\,{\bf 0}$
exponentially. Obviously, $\bar{{\bf Y}}(t)\,\rightarrow\,{\bf 0}$
exponentially is equivalent to ${\bf Y}(t)\,\rightarrow\,{\bf 0}$
exponentially, where ${\bf Y}(t)\,=\,\left({\bf y}_1(t),\,{\bf
y}_2(t),\,\cdots,\,{\bf y}_N(t)\right)\in\textbf{R}^{n\times N}$.

Let ${\bf Y}(t)\,=\,{\bf V}(t)\,{\bf\Phi}^{-1}(t)$, where ${\bf
\Phi}(t)$ is defined in the statements of Theorem 3. Similar to
the proof of Theorem 1, we have
\begin{equation}\label{23}
  \dot{{\bf v}}_k\,=\,[D{\bf f}({\bf s}(t))\,+\,\lambda_k(t)\,
                    {\bf A}(t)\,-\,\beta_k(t)\,{\bf I}_n]\,
                    {\bf v}_k\,,\quad\quad k\,=\,1,\,2,\,\cdots,\,N\,.
\end{equation}
Thus, we have changed the exponential stability problem of the
synchronous solution ${\bf S}(t)$ into the exponential stability
problem for $N$ independent $n-$dimensional linear systems
(\ref{23}), which has the same form as (\ref{18}).

From Assumption 1 and Lemma 1, $\lambda_1(t)\,=\,0$ for all
$t\,>\,t_0$, and so the corresponding system (\ref{23}) becomes
system (\ref{21}). From the assumed conditions of Theorem 3, ${\bf
v}_1(t)\,\rightarrow\,{\bf 0}$ exponentially. Moreover, ${\bf
Y}(t)\,\rightarrow\,{\bf 0}$ exponentially is equivalent to ${\bf
v}_k(t)\,\rightarrow\,{\bf 0}$ exponentially for
$k\,=\,2,\,\cdots,\,N$. Therefore, the synchronous solution ${\bf
S}(t)$ is exponentially stable for network (\ref{3}) if and only
if the linear time-varying systems (\ref{18}) are exponentially
stable about their zero solutions. The proof is thus completed.

\bigbreak \noindent{\bf Remarks:} Theorem 3 shows that
synchronization of the time-varying dynamical network (\ref{3}) is
completely determined by its inner-coupling matrix ${\bf A}(t)$,
and the eigenvalues $\lambda_k(t)\;(2\,\leq\,k\,\leq\,N)$ and the
corresponding eigenvectors ${\bf\phi}_k(t)$ ($\beta_k(t)$ are
functions of ${\bf\phi}_k(t)\,,\,2\,\leq\,k\,\leq\,N$) of the
coupling configuration matrix ${\bf C}(t)$. However, the
synchronization of the time-invariant dynamical network (\ref{4})
is completely determined only by its inner-coupling matrix ${\bf
A}$ and the eigenvalues of the coupling configuration matrix ${\bf
C}$ [14,15].

\subsection*{C. A Simulation Example}

In this subsection, we illustrate Theorem 3 by using a
$3$-dimensional exponentially stable system as a node in network
(\ref{3}). For simplicity, we only consider a three-node network.
Each individual node is described by $\dot{x}_{1}\,=\,-\,x_{1}$,
$\dot{x}_{2}\,=\,-2\,x_{2}$, $\dot{x}_{3}\,=\,-3\,x_{3}$, which is
exponentially stable at ${\bf s}(t)\,=\,{\bf 0}$, and its Jacobian
is $D{\bf f}({\bf x})\,=\,\mbox{diag}\{-1,\,-2,\,-3\}$.

Assume that the inner-coupling matrix is ${\bf
A}(t)\,=\,\mbox{diag}\{1\,+\,e^{-2t},\,1\,+\,e^{-3t},\,1\,+\,e^{-t}\}$.
and the coupling configuration matrix is
$$
    {\bf C}(t)\,=\,\frac{1}{2e^2\,-\,e\,-1}\left(\begin{array}{ccc}
      c_{11}(t) & c_{12}(t) & c_{13}(t) \\
      c_{21}(t) & c_{22}(t) & c_{23}(t) \\
      c_{31}(t) & c_{32}(t) & c_{33}(t)
    \end{array}\right)\,,
$$
where
$c_{11}(t)\,=\,(e^2\,-\,1)\,\mbox{th}(t)\,+\,e\,\arctan(t),\;
c_{12}(t)\,=\,(1\,-\,e)\,\mbox{th}(t)\,-\,2e\,\arctan(t),\;
c_{13}(t)\,=\,(e\,-\,e^2)\,\mbox{th}(t)\,+\,e\,\arctan(t),\;
c_{21}(t)\,=\,2(e^2\,-\,1)\,\mbox{th}(t)\,+\,e^2\,\arctan(t),\;
c_{22}(t)\,=\,2(1\,-\,e)\,\mbox{th}(t)\,-\,2e^2\,\arctan(t)$,
$c_{23}(t)\,=\,2(e\,-\,e^2)\,\mbox{th}(t)\,+\,e^2\,\arctan(t),\;
c_{31}(t)\,=\,3(e^2\,-\,1)\,\mbox{th}(t)\,+\,\arctan(t),\;
c_{32}(t)\,=\,3(1\,-\,e)\,\mbox{th}(t)\,-\,2\arctan(t),\;
c_{33}(t)\,=\,3(e\,-\,e^2)\,\mbox{th}(t)\,+\,\arctan(t)$, with
th$(t)\,=\,\frac{e^t\,-\,e^{-t}}{e^t\,+\,e^{-t}}$.

It is easy to verify that there exists a nonsingular real matrix,
$$
{\bf \Phi}(t)\,=\,\frac{1}{2e^2\,-\,e\,-1}\left(\begin{array}{ccc}
      3e^2\,-\,2 & (1\,-\,e^2)e^{t} & -e^{1\,+\,\sin(t)} \\
      1\,-\,3e & (e\,-\,1)e^{t} & 2e^{1+\sin(t)} \\
      2e\,-\,e^2 & (e^2\,-\,e)e^{t} & -e^{1\,+\,\sin(t)}
    \end{array}\right)\,,
$$
such that ${\bf \Phi}^{-1}(t)\,\left({\bf C}(t)\right)^T\,{\bf
\Phi}(t)\,=\,\mbox{diag}\{0,\,-\mbox{th}(t),\,-\arctan(t)\}$ and
$\dot{{\bf \Phi}}^{-1}(t)\,{\bf \Phi}(t)\,=\,\mbox{diag}\{0,\,-1,$\\
$-\cos(t)\}$. Obviously, the conditions of Theorem 3 are
satisfied. Therefore, the zero solution of network (\ref{3}) in
this example is exponentially stable if and only if the linear
time-varying systems
\begin{equation}\label{27}
  \dot{{\bf w}}\,=\,[D{\bf f}({\bf s}(t))\,+\,\lambda_k(t)\,{\bf A}(t)\,
                  -\,\beta_k(t)\,{\bf I}_3]\,{\bf w}\,,\quad\quad
                  k\,=\,2,\,3,
\end{equation}
are exponentially stable about their zero solutions.

When $k\,=\,2$, we have $\lambda_2(t)\,=\,-\mbox{th}(t)$,
$\beta_2(t)\,=\,-1$. Then, we have, for all $t\,\geq\,1$,
$\mu\,[D{\bf f}({\bf s}(t))+\lambda_k(t){\bf A}(t)-\beta_k(t){\bf
I}_3]\,=\,-\left(1+e^{-2t}\right)\,\mbox{th}(t)\,<\,-\mbox{th}(1)\,<\,0$,
showing that the decoupled system (\ref{27}) with $k=2$ is
exponentially stable about its zero solution.

When $k\,=\,3$, we have $\lambda_3(t)\,=\,-\arctan(t)$,
$\beta_3(t)\, =\,-\cos(t)$. Similarly, we can verify that system
(\ref{27}) with $k=3$ is also exponentially stable about its zero
solution.

Therefore, from Theorem 3, the synchronous solution ${\bf
S}(t)\,=\,{\bf 0}$ of network (\ref{3}) is exponentially stable.

Note that, in this example, ${\bf C}(t)$ is not a constant matrix,
as assumed in other investigations [7-12,15,17,21].

\section*{\center{IV. CONCLUSIONS}}

Over the last few years, the discovery of small-world and
scale-free properties of many complex dynamical networks has seen
great advances, while these networks all have constant and uniform
connections. In this paper, we have further introduced a
time-varying dynamical network model, and investigated its
synchronization criteria. Several network synchronization theorems
have been established for this model. Especially, we have shown
that: synchronization of a time-varying dynamical network is
completely determined by its inner-coupling matrix, and the
eigenvalues and the corresponding eigenvectors of the coupling
configuration matrix, differing from that for time-invariant
complex networks studied elsewhere before.

The proposed time-varying complex network model provides a
mathematical description for further investigating the dynamical
behaviors and topological structures of many real-world complex
dynamical networks. We foresee that this model will be very useful
for the current intensive studies of general complex dynamical
networks.

\bigskip
\begin{center}
\textbf{ACKNOWLEDGEMENT}
\end{center}

Dr. Jinhu L\"u was deeply indebted to Prof. Lei Guo for some
stimulating discussions and helpful comments on this subject of
study.

\section*{\centerline{REFERENCES}}

\noindent [1] S. H. Strogatz, ``Exploring complex networks,'' {\it
Nature}, vol. 410, pp. 268-276, 2001.

\noindent [2] R. Albert and A. -L. Barab\'asi, ``Statistical
mechanics of complex networks,'' {\it Rev. Modern Phys.}, vol. 74,
pp. 47-97, 2002.

\noindent [3] A. -L. Barab\'asi and R. Albert, ``Emergence of
scaling in random networks,'' {\it Science}, vol. 286, pp.
509-512, 1999.

\noindent [4] D. J. Watts and S. H. Strogatz, ``Collective
dynamics of small-world,'' {\it Nature}, vol. 393, pp. 440-442,
1998.

\noindent [5] P. M. Gade, ``Synchronization in coupled map
lattices with random nonlocal connectivity,'' {\it Phys. Rev. E},
vol. 54, no. 1, pp. 64-70, 1996.

\noindent [6] A. R. Volkovskii and N. F. Rulkov, ``Experimental
study of bifurcations on the threshold for stochastic locking,''
{\it Sov. Tech. Phys. Lett.}, vol. 15, pp. 249, 1989.

\noindent [7] J. M. Kowalski, G. L. Albert, and G. W. Gross,
``Asymptotic synchronous chaotic orbits in systems of excitable
elements,'' {\it Phys. Rev. A}, vol. 42, no. 10, pp. 6260-6263,
1990.

\noindent [8] N. Nakagawa and Y. Kuramoto, ``Collective chaos in a
population of globally coupled oscillators,'' {\it Progress of
Theoretical Physics}, vol. 89, no. 2, pp. 313-323, 1993.

\noindent [9] L. M. Pecora and T. L. Carroll, ``Synchronization in
chaotic systems,'' {\it Phys. Rev. Lett.}, vol. 64, no. 8, pp.
821-824, 1990.

\noindent [10] L. M. Pecora, T. L. Carroll, G. A. Johnson, D. J.
Mar, and J. F. Heagy, ``Fundamentals of synchronization in chaotic
systems, concepts, and applications,'' {\it Chaos}, vol. 7, no. 4,
pp. 520-543, 1997.

\noindent [11] C. W. Wu and L. O. Chua, ``A unified framework for
synchronization and control of dynamical systems,'' {\it Int. J.
Bifurcation Chaos}, vol. 4, no. 4, pp. 979-998, 1994.

\noindent [12] T. Yamada and H. Fujisaka, ``Stability theory of
synchronized motion in coupled-oscillator systems II,'' {\it
Progress in Theoretical Physics}, vol. 70, no. 5, pp. 1240-1248,
1983.

\noindent [13] L. O. Chua, {\it CNN: A Paradigm for Complexity}.
Singapore: World Scientific, 1998.

\noindent [14] X. Wang and G. Chen, ``Complex network:
Small-world, scale-free and beyond,'' {\it IEEE Circ. Syst.
Magazine}, vol. 3, no. 2, pp. 6-20, 2003.

\noindent [15] X. Wang and G. Chen, ``Synchronization in
scale-free dynamical networks: Robustness and Fragility,'' {\it
IEEE Trans. Circ. Syst. -I}, vol. 49, pp. 54-62, Jan. 2002.

\noindent [16] P. Erd\"os and A. R\'enyi, ``On the evolution of
random graphs,'' {\it Publ. Math. Inst. Hung. Acad. Sci.}, vol. 5,
pp. 17-60, 1959.

\noindent [17] J. L\"u, T. Zhou and S. Zhang, ``Chaos
synchronization between linearly coupled chaotic system,'' {\it
Chaos, Solitons and Fractals}, vol. 14, no. 4, pp. 529-541, 2002.

\noindent [18] J. L\"u, X. Yu and G. Chen, ``Chaos synchronization
of general complex dynamical networks,'' {\it Physica A}, vol.
334, no. 1-2, pp. 281-302, 2004.

\noindent [19] J. L\"u, X. Yu, G. Chen and D. Cheng,
``Characterizing the synchronizability of small-world dynamical
networks,'' {\it IEEE Trans. Circ. Syst. -I}, vol. 51, no. 4, pp.
787-796, Apr. 2004.

\noindent [20] A. Turing, ``The chemical basis of morphogenesis,''
{\it Philosophical Transactions of the Royal Society of London,
B}, vol. 237, pp. 37-72, 1952.

\noindent [21] C. W. Wu and L. O. Chua, ``Synchronization in an
array of linearly coupled dynamical systems,'' {\it IEEE Trans.
Circ. Syst. -I}, vol. 42, no. 8, pp. 430-447, Aug. 1995.

\noindent [22] H. Minc, {\it Nonnegative Matrices}. New York:
Wiley, 1988.

\noindent [23] W. A. Coppel, {\it Stability and Asymptotic
Behavior of Differential Equation}. Boston: D. C. Heath and
Company, 1965.

\noindent [24] H. K. Khalil, {\it Nonlinear Systems}. New Jersey:
Prentice-Hall Inc., 1996.
\end{document}